\documentstyle[12pt]{article}

\newcommand{\wh}{\widetilde H}

\begin{document}

\title {Neutral Kaons without  Weisskopf--Wigner
Approximation}
\date{}
\author{ Ya.I.Azimov\\
Petersburg Nuclear Physics Institute\\
Gatchina, St.Petersburg 188350, Russia\\
e-mail: AZIMOV@THD.PNPI.SPB.RU}
\maketitle

\begin{abstract}
 The model-independent formalism is constructed to describe
decays of mixed particles without using the Weisskopf-Wigner
approximation. Limitations due to various symmetries are traced for
neutral $K-$mesons. As an application we show that effects of
$CPT-$violation and going beyond WWA may be separated and studied
independently.
\end{abstract}

\section{Introduction}
Neutral kaons present the best-known example of rather rare
phenomenon, quantum interference of massive particles. The
corresponding interference effects allowed to do unique measurements.
We mean, first of all, discovery and investigation of the
$CP$-violation.

New data on neutral kaons, with higher statistics and better
precision, should appear in the near future. The main goals of their
using seem to be searches for the $CPT$-\-violation and the so called
direct $CP$-\-violation (see, e.g.,[1--3]). To reach those goals one
needs to measure very small quantities (present experimental bounds
for the direct $CP$-violation see in [4]; estimates for possible
$CPT$-\-violation may be found, e.g., in [5]). Such tiny effects may
be influenced by various  approximations (or going beyond them). In
particular, corrections to the conventional Weisskopf-\-Wigner
approximation (WWA) [6,7] might imitate some properties of neutral
kaons which are usually related to $CPT$-\-violation [8] (compare to
the final state interaction imitating $T$-\-violation in decays).

Deviations from WWA (i.e., in essence, from the exponential
$t$-dependence) could be interesting by themselves. Remember here,
that the familiar exponential behavior of decaying unstable systems
can be realized in quantum theory only approximately. It should
inevitably be violated for very  small or very large times [9].
Related theoretical problems have been repeatedly discussed in
literature (see, e.g., recent papers [10,11] and references therein
to earlier works). But experiments came to success [12] only for
small times (the quantum Zeno effect), effects at large times stay
unobserved. Neutral kaons, which decays produce oscillations
sensitive to various small effects, could appear useful here as well.

The present paper studies evolution of neutral kaons without using
WWA. The problem has been investigated in [13,14], and later in [10].
But the approach of Refs. [13,14] is saturated by mathematics, that
screens the underlying physical meaning. Ref.[10] uses a particular
toy model. Here we assume only the many-\-channel character of the
problem and existence of the Hamiltonian.  The necessary formalism is
first described for decays of usual one-\-particle states (Sections
2,3).  Then it is generalized to decays of mixed states (Section 4).
Section 5 concerns with specific features of $K^0$ and $\bar K^0$,
especially with various possible symmetries. As application, we
consider in Section 6 the possible effects of $CPT$-\-violation and
going beyond WWA for neutral kaon decays.

\section{Evolution and survival of states}

The Schr\"odinger equation implies that the wave function evolves in
time as
\begin{equation}
\Psi(t)=e^{-iHt}\Psi^{(0)},
\end{equation}
where $H$ is the total Hamiltonian of the system, $\Psi^{(0)}$ is its
wave function at $t=0$. Useful for description  of decays is the
survival amplitude [11,15] equal to
\begin{equation}
A(t)=\langle \Psi^{(0)}\mid\Psi(t)\rangle=\langle\Psi^{(0)}\mid
e^{-iHt}\mid\Psi^{(0)}\rangle.
\end{equation}
If we are interested in the evolution only for $t>0$, we can rewrite
it as
$$ e^{-iHt}\cdot\Theta(t)=\frac i{2\pi}\int^\infty_{-\infty} dE
e^{-iEt}\frac1{E-H+i\epsilon}, $$
\begin{equation}
A(t)=\frac i{2\pi}\int^\infty_{-\infty}dEe^{-iEt}\langle\Psi^{(0)}|
\frac1{E-H+i\epsilon}|\Psi^{(0)}\rangle.
\end{equation}
A stationary state having
$$ H\Psi^{(0)}=E^{(0)}\Psi^{(0)} $$
gives to $A(t)$ the simple time dependence of $\exp(-iE^{(0)}t)$ with
real $E^{(0)}$ and constant absolute value. Exponential decay would
require complex-\-valued $E^{(0)}$. It cannot be an eigen-\-value of
any hermitian Hamiltonian, but the corresponding state can be defined
in a special sense [11]. However, one is unable to present such a
state as a combination of usual physical states. Therefore, we prefer
another approach.

Note, first of all, that description of unstable particles uses, as a
rule, several kinds of interaction. These may be interactions of
quite different nature (e.g., weak decays of hadrons), or various
manifestations of the same interactions (as for light emission by the
excited atom). Hence, for simplicity we begin with a system of two
coupled channels having interactions both inside and between
the channels.

The wave function is now a two-component column containing wave
functions of those two channels; Hamiltonian is a $2\times2$ matrix
\begin{equation}
H\;=\left(\begin{array}{ll} H_{11}& H_{12}\\ H_{21} & H_{22}
\end{array}\right).
\end{equation}
Hermitian character of $H$ means
\begin{equation}
H^{\dag}_{11}=H_{11}, \qquad H^{\dag}_{22}=H_{22}, \qquad
H^{\dag}_{12}=H_{21}.
\end{equation}
We use here the symbol of hermitian, not complex, conjugation because
of the operator nature of the matrix elements $H_{jk}$. Each of the
indices, 1 and/or 2, may be considered actually as related to a set
of channels. Expressions (1)--(3) stay applicable to the
two-\-component wave function $\Psi(t)$ if one  takes in them the
matrix (4) for $H$.

Consider now a more special case where $\Psi^{(0)}$ has only one
nonvanishing component $\Psi^{(0)}_1$ corresponding to channel(s) 1.
Then we can simplify the survival amplitude. Indeed, only the
operator
\begin{equation}
R(E)=\left(\frac1{E-H}\right)_{11}
\end{equation}
without additional matrix structure is essential here. From the
matrix identity
\begin{equation}
(E-H)\cdot\left(\frac1{E-H}\right)=1
\end{equation}
we obtain:
\begin{equation}
R(E)=\frac1{E-\wh(E)},
\end{equation}
where
\begin{equation}
\wh(E)=H_{11}+H_{12}\cdot\frac1{E-H_{22}}\cdot H_{21}.
\end{equation}

Expression (8) exactly corresponds to expression (3) for the pure
one-\-channel problem. So $\wh$ appears as an effective one-\-channel
Hamiltonian. But the influence of other channels generates two
important effects.\\

1) Unlike the true Hamiltonian $H$, the effective Hamiltonian $\wh$
depends on $E$, thus revealing time "nonlocality" of interaction in
one channel because of transition to other channel(s) and return. In
fact, we can obtain $A(t)$ by using an effective one-component wave
function $\Psi_1(t)$ such that \begin{eqnarray} A(t)&=&
\langle\Psi^{(0)}_1\mid\Psi_1(t)\rangle, \nonumber\\
\Psi_1(t)\cdot\Theta(t)&=&\frac i{2\pi}\int^\infty_{-\infty}
dEe^{-iEt}R(E+i\epsilon)\Psi^{(0)}_1.
\end{eqnarray}
The so defined function $\Psi_1(t)$ is the first component of the
two-\-component wave function $\Psi(t)$ if
$$ \Psi(0) =\left( {\Psi^{(0)}_1 \atop 0} \right). $$
It satisfies the integro-differential equation
\begin{equation}
i\frac d{dt}\Psi_1(t)=\int^t_0 dt'H_1(t-t')\Psi_1(t'),
\end{equation}
where the lower limit appears because we study the evolution only for
$t>0$, the upper limit expresses causality. Here
\begin{equation}
H_1(t)=\frac1{2\pi}\int^\infty_{-\infty}dEe^{-iEt}\wh(E+i\epsilon).
\end{equation}

Equations (11),(12) demonstrate nonlocal character of time evolution
of $\Psi_1(t)$. Were $\wh$ independent of $E$, equation (11) would
transform into the familiar Schr\"odinger equation, local in time.\\

2) Even if the total Hamiltonian $H$ is hermitian, the effective
Hamiltonian $\wh$ may be nonhermitian. Due to relations (5), the
definition (9) implies
\begin{equation}
\wh^{\dag}(E) =\wh(E^*).
\end{equation}
Hence, $\wh$ becomes nonhermitian both for complex values of $E$ and
for real values lying at the spectrum, point-\-like or continuous, of
the operator $H_{22}$.

The above presentation shows that expressions (8),(9) are quite
exact. However, they coincide with formal summation of perturbation
series (i.e. expansion in $H_{12}$ and $H_{21}$). In particular, they
allow to obtain the standard perturbative expressions [16] for energy
eigen-\-values. It is interesting to note that due to the coupling of
channels one can use the same expression (9) as the basis to find
energies of states which, in the decoupling limit, would refer to
both channel 1 and channel 2.

Consider briefly eigen-states of the total Hamiltonian. Each of them
has the corresponding two-\-component wave function solving the
familiar eigen-\-value equation
\begin{equation}
H\Psi^{(n)}=E^{(n)}\Psi^{(n)},
\end{equation}
$H$ being the total Hamiltonian. One can easily check that its
channel 1 component $\Psi^{(n)}_1$, by itself, solves an unusual
eigen-\-value equation:
\begin{equation}
\wh(E^{(n)})\Psi^{(n)}_1=E^{(n)}\Psi_1^{(n)},
\end{equation}
with the same value $E^{(n)}$ as in Eq.(14). $E^{(n)}$ is, surely,
real because of hermiticity of $H$. The fact that the set of
$\Psi^{(n)}$ is complete and orthonormalized implies completeness of
the set of $\Psi_1^{(n)}$. Note, however, that $\Psi^{(n)}_1$ are not
orthonormalized. Equation (15) gives one more illustration of
$\wh(E)$ as an effective Hamiltonian in channel 1.

\section{Decays of a separate state}

Initial state $\Psi^{(0)}_1$ has not been fixed till now. Let us take
it to be an eigen-\-state of the operator $H_{11}$ having energy
$E^{(0)}_1$ and wave function $\psi^{(0)}_1$:
\begin{equation}
H_{11}\psi^{(0)}_1 =E^{(0)}_1\psi^{(0)}_1.
\end{equation}
Such a state would be stationary if transitions to channel(s) 2 were
absent. Due to the transitions it will spread. According to the
previous Section the survival amplitude of that state is related to
the matrix element
\begin{equation}
a(E)=\langle\psi^{(0)}_1\left|\frac1{E-\wh(E)}\right|
\psi^{(0)}_1\rangle.
\end{equation}
It can be presented as
\begin{equation}
a(E)=\frac1{E-h(E)}.
\end{equation}
$h(E)$ is a numerical function defined, in essence, by Eqs.(17),(18).
It can be written also in the form
\begin{equation}
h(E)=E^{(0)}_1+\frac1{a(E)}\langle\psi^{(0)}_1\left|
H_{12}(E-H_{22})^{-1}H_{21}(E-\wh)^{-1}\right|\psi^{(0)}_1\rangle
\end{equation}
convenient for perturbative expansion.

The function $h(E)$ has the same meaning for the state $\psi^{(0)}_1$
as $\wh(E)$ for the whole channel 1. So it can be considered as an
effective Hamiltonian of the state $\psi^{(0)}_1$, again nonlocal in
time because of $E$-\-dependence. Applying complex conjugation to
Eqs.(17),(18) we obtain relation
\begin{equation} h^*(E)=h(E^*) \end{equation}
directly analogous to (13). Similar to $\wh(E)$, the function $h(E)$
is complex even at real values of $E$ if they correspond to the
spectrum (point-\-like or continuous) of $H_{22}$.

Solution of the equation
\begin{equation}  E_1-h(E_1)=0  \end{equation}
produces the pole of $a(E)$ at $E=E_1$. If the solution is real, it
gives the eigen-\-value of the total Hamiltonian taking into account
interaction between channels 1 and 2. If $H_{12}$ and
$H_{21}=H^{\dag}_{12}$ contain a small parameter $\delta$ one can use
Eq.(21) to reconstruct standard perturbative formulas [16] for the
energy shift of the channel 1 state under influence of channel 2. But
if the value $E^{(0)}_1$ lies at the continuous spectrum of channel 2
(this is just decay case) then $h(E^{(0)}_1)$ and, therefore,
solution of Eq.(21) near $E^{(0)}_1$ are complex.

The survival amplitude of the state $\Psi^{(0)}_1$ equals
\begin{equation}
A(t)=\frac i{2\pi}\int^\infty_{-\infty} dEe^{-iEt}a(E+i\epsilon).
\end{equation}
For small $\delta$ the pole of integrand (22) appears at
\begin{equation}
E_1\approx h(E^{(0)}_1+i\epsilon).
\end{equation}
Then the structure of $\wh(E)$ leads to Im$\,E_1<0$, so the pole
produces exponential decrease of $A(t)$.

Now one can easily justify the necessity [9] of deviation of $A(t)$
from pure exponential behaviour at very small and very large $t$. For
small $t$ we can directly use Eq.(2) and have
\begin{equation}
A(0)=1, \qquad A(t)\approx1-it\langle\Psi^{(0)}|H|\Psi^{(0)}\rangle.
\end{equation}
Hermitian total Hamiltonian $H$ gives purely real value of
$\langle\Psi^{(0)}|H|\Psi^{(0)}\rangle$. Hence, the survival
probability, equal to $|A(t)|^2$, does not contain any term linear in
$t$. Its decrease from the initial unit value goes slower than
prescribed by the exponential.

More convenient for large $t$ are Eqs. (18),(22) which present $A(t)$
as the sum of contributions from singularities of $a(E)$. The closer
to the real axis is the singularity, the slower is exponential
decrease of its contribution. But $h(E)$ and $a(E)$ always contain
real singularities due to thresholds in various channels. It is just
the threshold contributions that give the slowest $t$ decrease. Each
of them goes as power law with the power exponent depending on
behavior of $h(E)$ near the threshold.

Return now to the small parameter $\delta$ in $H_{12}$ and $H_{21}$.
If perturbation expansion in $\delta$ is applicable, then only one
singularity (it is just the pole (21),(23)) produces
contribution without parametric smallness. All other contributions
have order $\sim \delta^2$ or higher. Thus, survival probability is
very close to the exponential and deviates only at very small or very
large $t$.  Note that the true decrease in both cases is slower than
exponential.

More diverse time behavior is possible if the perturbation expansion
in $\delta$ is inapplicable. E.g., two singularities can produce
contributions which are not small, but have opposite signs. At $t=0$
they may essentially subtract each other.  If one of them decreases
with $t$ faster than another, the survival probability may even
increase in some interval of $t$.

Generally, if various singularities give contributions of comparable
value then survival probability inevitably contains oscillating terms
due to interference of various contributions.

Now we are ready to formulate WWA for decay of a separate state [6].
To have it one should change
\begin{equation}
h(E)\to h(E^{(0)}_1+i\epsilon)
\end{equation}
under the integral (22). Eq.(23) for the pole position becomes exact;
integral (22) has only the pole contribution, and $A(t)$ appears to
be pure exponential of $t$. At first sight WWA means rejection of
contributions from all singularities of $a(E)$ except the pole. But
actually it changes normalization of the pole term as well, so to
reserve the  initial condition $A(0)=1$. Both effects have the same
order of smallness if perturbation theory is applicable.

To complete the Section we make one more note. The structure of
expression (18), with function of $E$ appearing instead of the
eigen-\-state energy, reminds the exact structure of one-\-particle
propagator in the quantum field theory where the mass operator
depending on the momentum replaces the mass value. One can easily
understand that the similarity is not accidental. Both phenomena have
the same reason. It is the influence of virtual transitions to other
channels.

\section{Decays of coupled states}

The above formalism is not convenient for neutral kaons and other
mixed systems, since their description requires to consider both
coupled channels at the same basis. That is why we somewhat modify
the approach.

Let us consider three-channel situation. Keeping kaons in mind we can
consider three-channel situation. Keeping kaons in mind we can take
channels 1 and 2 as having strangeness $S=\pm1$, while index
3 should correspond to the totality of channels  with all other
values of $S$. Wave function has now 3 components, Hamiltonian $H$
and resolvent $(E-H)^{-1}$ are $3\times3$ matrices. Matrix elements
of hermitian total Hamiltonian satisfy relations
\begin{equation}
H^{\dag}_{jk}=H_{kj} \qquad (j,k=1,2,3).
\end{equation}

Exclude channel 3 from explicit consideration, just as was done
earlier for channel 2. From the resolvent we separate its part
$\widetilde R(E)$ that describes evolution and mutual transitions of
channels 1 and 2. It can be expressed as
\begin{equation}
 \widetilde R(E)=(E-\wh)^{-1},
\end{equation}
similar to Eqs. (6),(8). But now the effective Hamiltonian $\wh$ is
two-\-channel; it is $2\times2$ matrix with elements $(j,k=1,2)$
\begin{equation}
\wh_{jk}(E)=H_{jk}+H_{j3}(E-H_{33})^{-1}H_{3k}.
\end{equation}
The two-channel effective Hamiltonian, just as for one-\-channel
case, depends on $E$, i.e. interaction is nonlocal in time. Hermitian
total Hamiltonian $H$ produces relations for $\wh$ (compare to
one-\-channel relation (13)):
\begin{equation}
\wh^{\dag}(E)=\wh(E^*). \end{equation}
Hence, the matrix operator $\wh(E)$ is hermitian only at real $E$
outside the spectral region of $H_{33}$.

We were specially interested earlier in evolution of one particular
state $\psi^{(0)}_1$. Now we will follow for two states,
$\psi^{(0)}_1$ in channel 1 and $\psi_2^{(0)}$ in channel 2.
Moreover, we will study both their survival amplitudes and their
mutual transitions. In other words, we consider now $A(t)$ as
$2\times2$ matrix with elements
\begin{equation}
A_{jk}(t)=\langle\psi^{(0)}_j\mid\Psi_k(t)\rangle, \qquad j,k=1,2.
\end{equation}
Here $\Psi_k(t)$ is the 3-component wave function with the initial
condition \begin{equation}  \Psi_k(0)=\psi^{(0)}_k, \end{equation} while
$\psi^{(0)}_j$ may
be viewed as having only one nonvanishing component which corresponds
to eigen-\-state of \begin{equation} H_{jj}\psi^{(0)}_j=E^{(0)}_j\psi^{(0)}_j.
\end{equation}
Then we obtain
\begin{equation}
A(t)=\frac i{2\pi}\int^\infty_{-\infty} dEe^{-iEt}a(E+i\epsilon),
\end{equation}
where $a(E)$ generalizes the quantity (17). It is now $2\times2$
matrix with elements
\begin{eqnarray}
a_{jk}(E) &=& \langle\psi^{(0)}_j\left|\frac1{E-\wh(E)}
\right|\psi^{(0)}_k\rangle \nonumber \\
&=& \langle\psi^{(0)}_j\left|\frac1{E-H}\right|\psi^{(0)}_k\rangle.
\end{eqnarray}
Similar to (18), the matrix $a(E)$ may be presented as
\begin{equation} a(E)=(E-h(E))^{-1}, \end{equation}
where $2\times2$ matrix $h(E)$ may be defined by the matrix relation
\begin{equation}
h(E)\cdot a(E)=\langle\wh(E)\widetilde R(E)\rangle.
\end{equation}
The symbol $\langle\ldots\rangle$ means "projecting" the product of
operators to 2-\-dimensional space spanned over the two
eigen-\-states. Eq.(36) is, surely, an analog of Eq.(19). And this is
the end of simple analogies. E.g., Eq.(21) cannot be readily
rewritten in the matrix form (one term would be multiple of the unit
matrix, another would not).

Nevertheless, one may consider $h(E)$ as an effective Hamiltonian of
two states, $\psi^{(0)}_1$ and $\psi^{(0)}_2$, just as earlier it was
an effective one-\-state Hamiltonian. By tracing relations
(27),(28),(34--(36), that connect $h(E)$ to the total Hamiltonian
$H$, one can verify that hermitian $H$ (see (26)) produces matrix
relation (compare to (29)):
\begin{equation}
h^{\dag}(E)=h(E^*).
\end{equation}

Consider in more detail $a(E)$ as an effective resolvent for two
coupled states. It is convenient to expand $h(E)$ over linearly
independent matrices:
\begin{equation}
h(E)=h_0(E)+\sum^3_{k=1}b_k(E)\sigma_k,
\end{equation}
where $\sigma_k$ are the Pauli matrices. New numerical functions
$h_0(E)$ and $b_k(E)$ are simply related to components of $h(E)$:
\begin{equation} \begin{array}{ll}
h_0=\frac12(h_{11}+h_{22}), \:&\: b_1=\frac12(h_{12}+h_{21}),\\
& \\
b_2=\frac i2(h_{12}-h_{21}), \;&\; b_3=\frac12(h_{11}-h_{22}).
 \end{array}   \end{equation}
If we denote eigen-values of $h(E)$ as $\lambda^{(1)}(E)$ and
$\lambda^{(2)}(E)$, then
\begin{eqnarray}
\lambda^{(1),(2)} & = &h_0\pm b, \nonumber \\
b&=&\sqrt{b^2_1 +b^2_2 +b^2_3 }.
\end{eqnarray}
Note that both $b_k(E)$ and $b(E)$ are complex. Therefore, the
problem arises how to formulate some additional condition so to fix
branch of the root in Eq.(40). We discuss it later.

Now we can present $h(E)$ as
\begin{equation}
h(E)=\lambda^{(1)}(E)\cdot P_+(E)+\lambda^{(2)}(E)\cdot P_-(E),
\end{equation}
where the projecting operators $P_{\pm}$ equal
\begin{equation}
P_{\pm}(E)=\frac{1\pm\vec n\vec \sigma}2, \qquad \vec n(E)=\frac{\vec
b(E)}{b(E)}.
\end{equation}
Generally, the vector $\vec n(E)$ has complex-\-valued components.
But it is a unit vector in the sense
\begin{equation}
\vec n^2=n^2_1+n^2_2+n^2_3=1.
\end{equation}
Respectively,
\begin{equation}
P_+P_-=P_-P_+=0\,,\qquad  P_++P_-=1\,, \qquad (P_{\pm})^2=P_{\pm}\,,
\end{equation}
but $P^{\dag}_{\pm}$ may not coincide with $P_{\pm}$.

Eq.(41) separates $h(E)$ to two parts and allows to do the same for
$a(E)$:
\begin{equation}
a(E)=\frac1{E-\lambda^{(1)}(E)}P_+(E)+\frac1{E-\lambda^{(2)}(E)}
P_-(E).
\end{equation}
Each of the two parts in (45) reminds the case of one-\-state decay
(see Eq.(18)). Two-\-state character, i.e. mixing and mutual
transitions of two decaying states, reveals itself "only" in
multiplication by projecting operators. As a result, the quantity
$a(E)$, being described in one-\-state case by a single complex
numerical function $h(E)$, requires for two-\-state case to specify
four, not two complex functions. They may be either 4 elements of
$2\times2$ matrix $h(E)$ (see Eq.(35)), or 2 functions
$\lambda^{(1)}(E)$, $\lambda^{(2)}(E)$ and 2 independent components
of the complex unit vector $\vec n(E)$ (see Eqs. (42),(43)) contained
in $P_{\pm}$ (see Eqs.(45),(42)). Surely, such complication is
directly related to the fact that for the two-\-state case we are
interested not only in survival amplitudes of the two states, but
also in amplitude of their mutual transitions.

Operators $P_{\pm}$ are orthogonal to each other (in the sense of the
two first Eqs.(44)) and, at first sight, should select two chosen
combinations in the 2-\-dimensional set of states. However,
$E$-dependence of $P_{\pm}$ makes the selection nonlocal in time and
physically inefficient.

Note one more feature of the two-state problem. If transition
elements of $H$ can be considered as perturbation then
$\lambda^{(1)}$  and $\lambda^{(2)}$ are expandable. But components
of $\vec n$ become nonexpandable if nonperturbative  spectra of
channels 1 and 2 coincide. It is manifestation of the well-\-known
property of quantum theory [16]: if a system contains
energy-\-degenerate states then perturbation theory can be applicable
only when the initial states have been changed by specially
(nonperturbatively) selected combinations of them. It is just this
nonperturbativeness that gives possibility for  survival amplitudes
to be oscillating (see discussion in the preceding Section).

Let us discuss some details of the structure of $A(t)$ for two-\-state
case. Its large $t$ behavior is determined now by two sets of
singularities in accordance with two-\-term structure (45). Among
them there are two poles, at $E^{(1)}$ and $E^{(2)}$, solving
equations
\begin{equation}
E^{(1)}-\lambda^{(1)}(E^{(1)})=0, \qquad E^{(2)}-\lambda^{(2)}
(E^{(2)})=0.
\end{equation}
They generate two different exponentials in $A(t)$ and, generally,
oscillations in survival probabilities. There exist, however, two
such combinations of initial states $\psi^{(0)}_1$ and $\psi^{(0)}_2$
which evolution does not reveal one or another of the exponentials.
If the initial two-\-component function $\psi^{(2)}(0)$ satisfies the
condition
\begin{equation}
P_+(E^{(1)})\psi^{(2)}(0)=0,
\end{equation}
then it will evolve without the exponential generated by the pole of
the first term of (45). Another initial state, $\psi^{(1)}(0)$ with
condition
\begin{equation}
P_-(E^{(2)})\psi^{(1)}(0)=0,
\end{equation}
will evolve without contribution from the pole of the second term of
(45).

Retaining in the integral (33) only pole contributions from both
terms of (45) one obtains the pole approximation suggested in [8].
The standard form of generalized WWA suggested in [7] to describe
neutral kaons appears if one retains the pole contributions and
neglects difference between $\vec n(E^{(1)})$ and $\vec n(E^{(2)})$.
Taking this difference into account is of higher order smallness if
one applies the perturbative expansion with respect to transition
elements of $H$.

\section{Decays of kaons}

Since $K^0$ and $\bar K^0$ can turn to each other their description
without WWA should use formalism of the preceding Section. But some
problems need to be considered in more detail.

We identify channel 1 with the set of states having strangeness
$S=+1$ while channel 2 corresponds to the set of states with $S=-1$.
Channel 3 includes all states with $S\neq\pm1$; the most important in
the framework of perturbation theory are states with $S=0$. Thus, we
may enumerate Hamiltonian matrix elements by the corresponding
strangeness values $(H_{1,-1}$ instead of $H_{12}$, $H_{10}$ instead
$H_{13}$ and so on). Further, we use $K^0$ and $\bar K^0$ as
$\psi^{(0)}_1$ and $\psi^{(0)}_2$.
Channels of various strangeness are coupled by weak interaction
which strongly violates space reflection $P$ and charge conjugation
$C$. Therefore, we do not consider them separately, but only their
combination $CP$. In the limit of $CP$-invariance we can use
$CP$-transformation to relate states of $S=+1$ and $S=-1$.
In particular, we define phases of one-\-particle states so that
\begin{equation}
\bar K^0=(CP)K^0.
\end{equation}
The set of states with $S\neq \pm1$ goes to itself  under
$CP$-transformation.

With these conventions we rewrite relations (26) for hermitian
Hamiltonian as
\begin{equation}
H^{\dag}_{jk}=H_{kj} \qquad (j,k=-1,0,+1).
\end{equation}
Various symmetries produce additional relations between elements of
the Hamiltonian. E.g., $CP$-\-invariance gives
\begin{equation}
\begin{array}{ll} H_{11}=H_{-1,-1}\,; & H_{1,-1}=H_{-1,1}\,; \\
 & \\
H_{10}=H_{-1,0}\,; & H_{0,1}=H_{0,-1}\, . \end{array}
\end{equation}
We can add time inversion $T$ and consider $CPT$-\-transformation.
Assumption of $CPT$-\-invariance gives smaller number of relations:
\begin{equation}
H_{11}=H_{-1,-1}\,; \qquad H_{10}=H^{\dag}_{-1,0}=H_{0,-1}\,.
\end{equation}
$CPT$-invariance does not produce any relations for $H_{1,-1}$ and
$H_{-1,1}$ additional to (50). Hermitian conjugation arises in (52)
due to antiunitary nature of transformations $T$ and $CPT$.

Consider now the effective two-channel Hamiltonian $\wh(E)$.
Hermitian $H$ leads to relations (29) that connect matrix elements of
$\wh$ at different values of $E$. Unlike those, symmetry relations
connect various elements of $\wh$ with the same $E$. So,
$CPT$-\-invariance gives:
\begin{equation}
\wh_{11}(E)=\wh_{-1,-1}(E).   \end{equation}
The same relation is true for $CP$-invariance which adds one more
relation:
\begin{equation}
\wh_{1,-1}(E)=\wh_{-1,1}(E).  \end{equation}
Note that $T$-invariance by itself also leads to Eq.(54) but does not
require Eq.(53).

Tracing connections of $h(E)$ and $\wh(E)$ we find that
$CPT$-\-invariance leads to
\begin{equation}
h_{KK}(E)=h_{\bar K\bar K}(E) \end{equation}
in analogy with (53). Similarly, $CP$-invariance gives Eq.(55) with
addition
\begin{equation}
h_{K\bar K}(E)=h_{\bar KK}(E),     \end{equation}
which is an evident analog of Eq.(54).

These relations take quite simple form for vectors $\vec b(E)$ and
$\vec n(E)$ (see Eqs. (39)--(43)). From $CPT$-\-invariance (55) we
obtain
\begin{equation} n_3(E)=0, \end{equation}
while $CP$-invariance relations (55),(56) give
\begin{equation} n_2(E)=n_3(E)=0.    \end{equation}
$T$-invariance requires only vanishing $n_2(E)$.

Note that all the symmetries do not influence $\lambda^{(1)}(E)$ and
$\lambda^{(2)}(E)$ but essentially change properties of $\vec n(E)$.
For $CP$-\-invariant case there is only one non-\-zero component of
$\vec n$ which looses any $E$-dependence due to normalization (43).
$CPT$-\-invariance does not eliminate $E$-dependence but implies only
one independent component of $\vec n$ (two non-\-vanishing components
are related to each other by normalization).

Existing data imply that $CP$-violation is small while
$CPT$-\-violation has not been observed at all. This means
\begin{equation}
|n_3(E)|\ll|n_2(E)|\ll|n_1(E)|     \end{equation}
(note that we really know it only for $E\approx m_K$). Therefore, we
can choose the branch of root in Eq.(40) so as
\begin{equation}
b(E)\approx b_1(E), \qquad n_1(E)\approx1-\frac12 n^2_2(E).
\end{equation}
Further, when applying to kaons we denote $\lambda_S$ and $\lambda_L$
instead of $\lambda^{(1)}$ and $\lambda^{(2)}$ respectively.

Solutions of the equations
\begin{equation}
M_S=\lambda_S(M_S), \qquad M_L=\lambda_L(M_L)     \end{equation}
give positions of two poles in $a(E)$ that generate exponential
contributions to $A(t)$. In analogy with $\psi^{(1)}$ and
$\psi^{(2)}$ (see Eqs. (47),(48)), there exist two initial states,
$K_S$ and $K_L$, which evolutions reveal only one exponential from
two possible ones. If we describe their (anti)kaon content by the
standard parameters $\epsilon_S$ and $\epsilon_L$ and apply relations
corresponding to Eqs. (47) and (48) then
\begin{eqnarray}
\frac{1-\epsilon_L}{1+\epsilon_L}&=&\left.\frac{n_1+in_2}{1-n_3}
\right|_{E=M_S}\,; \nonumber\\   & &\\
\frac{1-\epsilon_S}{1+\epsilon_S}&=&\left.\frac{n_1+in_2}{1+n_3}
\right|_{E=M_L}\,. \nonumber
\end{eqnarray}
We see that without WWA
\begin{equation} \epsilon_S\neq\epsilon_L \end{equation}
even in the case of $CPT$-invariance (i.e. at $n_3=0$) [8].
Meanwhile, it is just the relation (63) which is usually considered
to be manifestation of $CPT$-\-violation.

Note, however, that the states $K_S$ and $K_L$ as defined above
appear to have  unfamiliar properties. Evolution of each of them
reveals only one exponential term indeed. But there are power law
terms as well. As a result, the kaon content of the states changes
with time. In this sense one may say that the states $K_S$ and $K_L$
are not independent beyond WWA and regenerate each other [13,14].

Surely, all the construction becomes quite standard in the framework
of WWA when one discards nonexponential terms and $E$-dependence of
$\vec n$. Note, however, that this dependence  becomes eliminated
also by $CP$-\-invariance when
\begin{equation}
n_2=n_3=0, \qquad n_1=1. \end{equation}
In such a case the states $K_S=K_1$ and $K_L=K_2$ appear to be
totally decoupled, so their content does not change with time even
without WWA.

\section{Effects of violation of $CPT$ and WWA}

Possibility to imitate $CPT$-violation beyond WWA requires to study
in more detail the corresponding effects and their mutual influence.
Simultaneously we can have a simple illustration of how the above
formalism works.

We begin with the matrix $A(t)$. Its elements describe two kinds of
processes. One of them is survival of (anti)kaons, i.e. two
transitions: $K^0\to K^0$, $\bar K^0\to\bar K^0$. Their amplitudes,
according to Eqs.(33),(42),(45), are equal
\begin{eqnarray}
A_{KK}(t)&=&\frac
i{4\pi}\int\limits^\infty_{-\infty}dEe^{-iEt}[f_+(E+i\epsilon)+
n_3(E)f_-(E+i\epsilon)], \nonumber \\  &&\\
A_{\bar K\bar K}(t)&=&\frac
i{4\pi}\int\limits^\infty_{-\infty}dEe^{-iEt}[f_+(E+i\epsilon)-
n_3(E)f_-(E+i\epsilon)], \nonumber
\end{eqnarray}
where
\begin{equation}
f_{\pm}(E)=\frac1{E-\lambda_S(E)}\pm\frac1{E-\lambda_L(E)}.
\end{equation}
Other processes may be called inversion of kaons. They are
transitions $K^0\to\bar K^0$ and $\bar K^0\to\bar K^0$ with
amplitudes
\begin{eqnarray}
A_{\bar KK}(t)&=&\frac i{4\pi}\int\limits^\infty_{-\infty} dE
e^{-iEt}[n_1(E)+in_2(E)]f_-(E+i\epsilon), \nonumber\\
	       && \\
A_{K\bar K}(t)&=&\frac i{4\pi}\int\limits^\infty_{-\infty} dE
e^{-iEt}[n_1(E)-in_2(E)]f_-(E+i\epsilon), \nonumber
\end{eqnarray}
Amplitudes (65) contain $n_3(E)$ and are directly related to the
problem of $CPT$-\-invariance. The ratio of the same-species
amplitudes
\begin{equation}
R_{\rm same}\;=\;\frac{A_{KK}(t)}{A_{\bar K\bar K}(t)}
\end{equation}
deviates from unity if $CPT$ is violated. At the same
time it gains $t$-dependence. These properties are valid both in WWA
and beyond it.  The amplitudes (65) and their ratio (68) are
insensitive to violation or conservation of $CP$ if $CPT$ is
conserved.

Inversion amplitudes (67) have different properties. They are
influenced by conservation or violation of $CPT$ since $n_1$ and
$n_2$ are related to $n_3$ by normalization (43); but this effect is
of higher order smallness. However, it is just their ratio
\begin{equation}
R_{\rm opposite}=\frac{A_{\bar KK}(t)}{A_{K\bar K})t)}
\end{equation}
that demonstrates the so called $CP$-violation in kaon mixing. In the
framework of WWA
\begin{equation}
(R_{\rm opposite})_{\rm WWA}=\frac{n_1+in_2}{n_1-in_2}
=\frac{1-\epsilon_S}{1+\epsilon_S}\cdot\frac{1-\epsilon_L}{1+\epsilon_L}.
\end{equation}
If $CP$ is violated the ratio (69) becomes $t$-dependent beyond WWA
independently of conservation or violation of $CPT$. Note that
$t$-dependence of $R_{\rm opposite}$ is directly related to
$E$-dependence of $\vec n$. Therefore, this effect of going beyond
WWA has no analogs in decays of uncoupled states. We emphasize,
however, that it can be revealed only if $CP$-invariance (or, more
exactly, $T$-invariance) is violated.

Thus, $R_{\rm same}$ and $R_{\rm opposite}$ have essentially
different properties: $R_{\rm same}$ can depend on $t$ only at
violated $CPT$; on the other hand, $R_{\rm opposite}$  can depend on
$t$ only beyond WWA. These examples demonstrate that effects of
violation of $CPT$ or WWA can be discriminated and studied separately.
Unfortunately, quantitative theoretical estimates of both effects can
be made at present only in model-\-dependent ways.

\section{Conclusion}

Here we briefly summarize the above results. To describe kaon
evolution without Weisskopf-\-Wigner approximation we suggest
formalism which reminds modified one-\-particle propagator where the
mass is changed by the mass operator. Such approach makes more
transparent many results obtained earlier [13,14] in a highly
mathematical way. Symmetry considerations show, in particular, that
violation of $CPT$ and deviation from WWA can be studied
independently of each other.

The author acknowledges useful discussions with I.Bigi, R.M.Ryndin
and L.A.Khalfin.

The work was partly supported by the International Scientific
Foundation (grants NO~7000 and NO~7300).


\newpage
\begin{thebibliography}{99} \fussy

\bibitem{}  C.O.Dib, R.D.Peccei, Phys.Rev., {\bf D46}, 2265
(1992).
\bibitem{} M.Hayakawa, A.I.Sanda, Phys.Rev., {\bf D48}, 1150
(1993).
\bibitem{}  B.Winstein, L.Wolfenstein, Rev.Mod.Phys., {\bf 65},
1113 (1993).
\bibitem{}  Particle Data Group, Phys.Rev., {\bf D50},
No. 3, Pt.1 (1994).
\bibitem{}  V.A.Kostelecky, R.Potting, Phys.Rev., {\bf D51},
3923 (1995).
\bibitem{}  V.F.Weisskopf, E.P.Wigner, Zeit.f.Phys.,
{\bf 63}, 54 (1930); {\bf 65}, 18 (1930).
\bibitem{}  T.D.Lee, R.Oehme, C.N.Yang, Phys.Rev., {\bf 106},
340 (1957).
\bibitem{} Ya.I.Azimov, JETP Lett. {\bf 58}, 159 (1993).
\bibitem{} L.A.Khalfin, ZhETF {\bf33}, 1371 (1957).
\bibitem{}  C.B.Chiu, E.C.G.Sudarshan, Phys.Rev. {\bf D42},
3712 (1990).
\bibitem{} E.C.G.Sudarshan, C.B.Chiu, Phys.Rev., {\bf D47}, 2602 (1993).
\bibitem{} W.M.Itano et al., Phys.Rev.,
{\bf A41}, 2295 (1990).
\bibitem{}  L.A.Khalfin, in: "Group Theoretical Methods in Physics"
(Proc.of the 3rd Seminar, Yurmala, May 1985), v.2, M., "Nauka", 1986,
p.608;\\ L.A.Khalfin, Preprint of Leningrad Division of the Steklov
Mathematical Institute P-4-80, Leningrad, 1980 (in Russian).
\bibitem{}  L.A.Khalfin, Preprint DOE-ER 40200211, February 1990.
\bibitem{}  C.B.Chiu, B.Misra, E.C.G.Sudarshan, Phys.Rev.,
{\bf D16}, 520 (1977).
\bibitem{} L.D.Landau, E.M.Lifshitz, Quantum Mechanics, Fizmatgiz
Pub.House, 1963.
\end{thebibliography}
\end{document}